\documentclass[12pt]{article}
 \usepackage{amsfonts}
 \usepackage{amssymb}
 \usepackage{amsmath}
 \input epsf.tex
= 0cm \textwidth = 16cm \textheight = 24.5cm \topskip = -1cm
\newcommand{\be}{\begin{equation}}
\newcommand{\ee}{\end{equation}}
\newcommand{\bea}{\begin{eqnarray}}
\newcommand{\eea}{\end{eqnarray}}

\begin{document}
\title{Bound State of an Electron on a $^4$He Superfluid Droplet
 as a Test for Quantum Gravity}
\author{M. Haghighat\thanks{email: mansour@cc.iut.ac.ir}\ \ and
 \ \ F. Loran\thanks{email: loran@cc.iut.ac.ir}\\ \\
{\it Department of  Physics, Isfahan University of Technology}\\
{\it Isfahan,  Iran.} \\}

\date{}

\maketitle
\begin{abstract}
We address the problem of finding a system in  which there would
be measurable quantum gravitational effects. Following standard
quantum-field methods, we have calculated the first-order
radiative correction of graviton exchange on the binding energy
of  an electron with an ultra cold superfluid droplet of $^4$He
with mass about the Planck mass. For two $^4$He droplets with a 
mass difference of about one microgram, we show that the relative
difference in the binding energies is about one percent.
\end{abstract}
The most important obstacle in the search for quantum gravity
(QG) is the absence of the experimental input that is needed to
test the theory. There are papers on the phenomenology of the
quantum gravity \cite{pheno} but the magnitude of the proposed
QG-effects are expected to be negligible in standard
circumstances because they are suppressed by the Planck scale.
However, small effects can become observable in special contexts
and one can always search for an experimental setup such that a
very large number of the very small quantum-gravity contributions
are effectively summed together.
 For instance, the ratio of the
forces of gravity and electricity between two particles with mass
$m$ and charge  $e$ is
\begin{equation}
 \frac{F_g}{F_e}=\frac{Gm^2}{e^2}=\frac{\hbar c}{e^2}\left(\frac{m}{M_{\rm Plank}}\right)^2,
 \label{comparison}
 \end{equation}
 that is too small for the fundamental particles with the
 ordinary masses.
 It is well known that electrons can be bound in surface states
outside certain bulk dielectrics such as $^4$He \cite{pred}.
Experimentally, the binding energy of the electron to the surface
of liquid helium has been measured using millimeter-wave
spectroscopy to be 8 Kelvin \cite{8Kelvin}. R. Y. Chiao by
replacing the oil of the classic "Millikan oil drops" with
superfluid helium ($^4$He) with a gravitational mass of around
the Planck-mass scale ($M_{\rm Plank}\approx 22 $ micrograms ),
proposed an experimental setup for a gravity-wave antenna
\cite{chao}. In this Letter we propose another experiment to
detect the interplay between gravity and QED by exploring the
bound state of an electron to an ultra cold superfluid
 drop of $^4$He ($e$-$^4$He bound state).

 The $e$-$^4$He bound state is due to formation of image charge inside the
 dielectric sphere of the drop. The image potential is of
 the form \cite{book}
 \be
 V= -Ze^2/r,
 \label{potential}
 \ee
 where $r$ is radial distance of electron from the droplet
 surface and
 \be
 Z=\frac{(k-1)}{(k+1)},
 \label{Z}
 \ee
 in which $k\approx 1.057$ is the relative permittivity of
 Helium. Because there is a barrier of 1 eV, the penetration of the electron
 with binding energy of a few Kelvin, into the Helium drop is negligible.
 Thus, one can assume the $e$-$^4$He bound state as an
 ideal Hydrogen-like system. Using Eq.(\ref{Z}), one can see
 that the binding energy is about 8 Kelvin ($\approx 10^{-3}$ eV)  and
 the effective Bohr radius $r_{\rm B}=76\AA$ \cite{Grimes}.  This
 justifies the tick-film approximation we have considered to obtain
 Eq.(\ref{Z}) as far as the radius of the Helium drop is
 about one millimeter.

\begin{figure}[t]
 \centerline{\epsfxsize=2in\epsffile{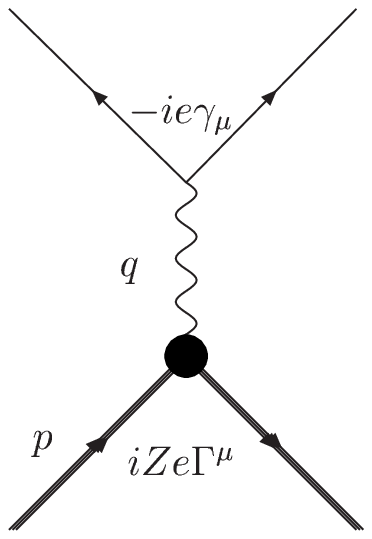}} \caption{
 One photon exchange diagram for $e$-$^4$He bound state. The upper legs show the
 electronic current, while the lower legs stand for the $^4$He drop current. }
 \label{interaction}
\end{figure}

\begin{figure}[t]
\centerline{\epsfxsize=6in\epsffile{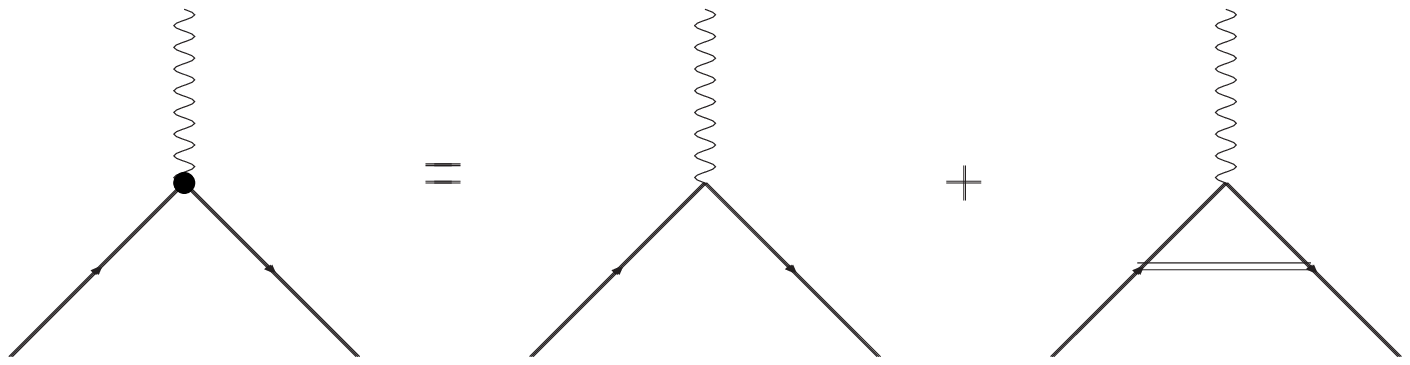}} \caption{ The
most significant gravitational one-loop correction for the
$^4$He-photon vertex. Double line represents graviton. }
\label{vertex}
\end{figure}

 In QED, the tree-level diagram corresponding to the
 hydrogen-like bound state is give by Fig.(\ref{interaction}).
In this letter we would like to explore the QG-effects in the
$e$-$^4$He bound state.  As is seen from Eq.(\ref{comparison}),
classically, only for particles with mass comparable to the Planck
mass the gravitational force is comparable with the electrical
force. Therefore, at tree level, the one graviton exchange
diagram between the electron and $^4$He is negligible in
comparison with the corresponding one photon exchange diagram. In
contrast, one anticipates that for the one loop corrections, the
QG-correction for the photon-$^4$He vertex to be the only
significant term in comparison with the other one loops coming
from both QED and QG, see Fig.(\ref{vertex}). Thus, gravitational
quantum corrections can have a valuable contribution to the
binding energy of $e$-$^4$He bound state. Consequently, this
mesoscopic system seems to be a good lab for probing the nature
of the interplay between gravitation and quantum mechanics.

The straightforward framework for general relativity quantized for
small fluctuations is a nonrenormalizable quantum field theory.
In fact, it is shown that the one-loop divergences in the theory
of scalar field coupled to quantum general relativity, can only
be cancelled by counterterms that do not appear in the original
action \cite{thooft}. Nevertheless, this framework is appropriate
for describing interactions at energies and momenta below the
Planck scale $M_{\rm Planck}$ when treated as an effective
low-energy theory \cite{Donoghue,Wilczeck}.

To calculate the one-loop diagram in Fig.(\ref{vertex}) we
consider the action for  transverse-traceless small fluctuation
modes $h_{\mu\nu}$, given by the following Lagrangian,
 \be
 {\cal L}=\frac{M_{\rm Planck}^2}{128\pi}
 \eta^{\alpha\beta}\partial_\alpha h^{\mu\nu}\partial_\beta
 h_{\mu\nu}+\left(\eta^{\mu\nu}+h^{\mu\nu}\right)
 \partial_\mu\phi^*\partial_\nu\phi+\cdots,
 \label{action}
 \ee
 in which $\eta_{\alpha\beta}$ is the Minkowski metric. In this
 way, for the low momentum transfer, one can show that
 \be
 \Gamma^\mu=p^\mu+\frac{1}{\pi}\frac{m^2_{^4{\rm
 He}}}{M_{\rm Planck}^2}\left\{-\frac{11}{6}+\frac{1}{2}\ln\left(\frac{\Lambda}{m_{^4{\rm
 He}}}\right)-\ln\left(\frac{\lambda}{2m_{^4{\rm
 He}}}\right)+{\cal O}\left(\frac{q^2}{m_{^4\rm
 He}^2}\right)\right\}p^\mu,
  \label{ras}
 \ee
 Where $\Lambda\sim M_{\rm Planck}$ is the ultraviolet cutoff and $\lambda$ is
 an infrared cutoff.  It should be noted that the mass of the helium droplet
 is not a fundamental parameter of the
quantum gravity and therefore the first three terms due to the
one loop correction in Eq.(5) can not be absorbed into any
properly renormalized vertex function. In fact the QG-structural
effects can be described by a form factor which, in principle, can
be measured in the e-$^4$He scattering processes for different
$^4$He droplets. The form factor at vanishing four momentum
transfer for an appropriate {\it reference mass} $m_r$ can be set
to unity.  Therefore the vertex function can be written as follows
 \be
 \Gamma^\mu=p^\mu+\frac{1}{\pi}\frac{m^2_{^4{\rm
 He}}}{M_{\rm Planck}^2}\left\{\ln\left(\frac{m_{^4{\rm
 He}}}{m_r}\right)+{\cal O}\left(\frac{q^2}{m_{^4\rm
 He}^2}\right)\right\}p^\mu,
  \label{ras2}
 \ee
 From Eq.(\ref{ras2}) one verifies that the  binding energy depends
 on the mass of $^4$He droplet. Therefore, the physical observable
 can be the difference in the binding energies  of different $e$-$^4$He droplets
 with different masses,
 \be
 \frac{\Delta E}{E}\simeq \frac{1}{\pi M^2_{\rm
 Planck}}\Delta \left(m^2_{^4\rm He}\ln{\frac{m_{^4\rm He}}{m_r}}\right).
 \ee
 where $E$ is the binding energy of $e$-$^4$He system at the tree level
 and  $\Delta E$ is the energy difference due to the one loop
 QG-correction for two $^4$He droplets with different
 masses.
 For instance, for two  $^4$He droplets with
 mass difference of about one microgram, and $m_r\simeq M_{\rm Planck}$ the
 relative difference in the binding energies will be about two percent.

\end{document}